\documentstyle[preprint,aps]{revtex}
\input epsf.tex
\def\DESepsf(#1 width #2){\epsfxsize=#2 \epsfbox{#1}}
\begin{document}

\preprint{\vbox{\hbox{OITS-617}\hbox{OSURN-318}\hbox{UTEXAS-HEP-96-18}
\hbox{DOE-ER 40757-087}}}
\draft
\title {Texture of fermion mass matrices in partially unified theories}
\author{B. Dutta$^{\dagger} $ and S. Nandi
$^{ \dagger \dagger} $ }\address{$^{\dagger} $ Institute of Theoretical Science,
University of Oregon, Eugene, OR 97403\\$^{\dagger \dagger} $ Department of
Physics,Oklahoma State University, Stillwater, Ok 74078\\and\\ Department of
Physics, University of Texas, Austin, Tx 78712\footnote{Present address, on
sabatical leave at the University of Texas at Austin.}}
\date{October, 1996}
\maketitle
\begin{abstract}  We investigate the texture of fermion mass matrices in
theories with partial unification (for example $ SU(2)_L\times SU(2)_R\times
SU(4)_c$) at a scale $\sim 10^{12}$ GeV. Starting with the low energy values of
the masses and the mixing angles, we find only two viable textures with atmost
four texture zeros. One of these corresponds to a somewhat modified Fritzsch
textures. A theoretical derivataion of these textures leads to new interesting
relations among the masses and the mixing angles.
\end{abstract}

\newpage Since the existence of a grand unifying scale is always well motivated,
several attempts have been made to unify the $SU(3)_c\times SU(2)_L\times
U(1)_Y$ in bigger groups e.g SU(5), SO(10), E$_6$ etc. But it is not decisive
yet which one among them is "the grand unfying group". Also it is not clear
whether there is no other scale, with some kind of symmetry being broken, in
between the weak scale and the grand unifying scale. There can always be a
partial unification happenning in between the two scales, since the  existence
of a partially unifying scale around
$10^{10}-10^{12}$ GeV in the SUSY theories is also well motivated for different
reasons. For example, one can naturally get a neutrino mass in the interesting
region of 3-10 eV, which could serve as hot dark matter candidate. Such a
neutrino mass may also be needed to explain the large scale structure of the
universe \cite{[SS]}. The window,
$10^{10}-10^{12}$ GeV, is also the right size for a hypothetical PQ symmetry to
be broken so as to solve the strong CP problem without creating any
phenomenological or cosmological problem \cite{[axion]}. This scale also gives
rise to lepton flavor violating prcesses like,
$\mu\rightarrow e\gamma$ \cite{[ddk1]} and electric dipole moment of electron and
neutron
\cite{[ddk3]} which may make it possible to detect this scale in the near future.
These lepton flavor violations and the edms get generated in these models without
having to make any assumption regarding the location of the soft SUSY breaking
terms. They can be anywhrere between the GUT scale and the Planck or string
scale. The GUT scale (with MSSM being the symmetry below the GUT scale) is lower
than the string scale by a factor of 25. However using the intermediate scale
around
$10^{12}$ GeV, it is possible to push up the GUT scale to the string scale
\cite{[ddk2]}. Also recently , an intermediate scale at $\sim 10^{12} $ GeV has
been advocated \cite{[TK]} to produce monopoles that would explain the high
energy cosmic ray spectrum.  The partially unified models like $SU(2)_L\times
SU(2)_R\times SU(4)_C$ gives rise to such monopoles naturally.

It is always a big challange to find a justification for the the mass
distribution of the fermions. So far attempts have been made from the GUT scale
and the weak scale \cite{[frit],[SOTM],[na]}. In this letter we try to generate
the most predictive textures with minimum number of parameters at a scale $\sim
10^{12}$ GeV where the theory is partially unified. We do not use any effect of
any particular grand unifying group, however for the partial unification group
we use
$SU(2)_L\times SU(2)_R\times SU(4)_C$ symmetry. We write down the superpotential
for these mass matrices and also we write some new relations among the masses and
the mixing angles.

We use the bottom up approach as developed in the reference \cite{[na]} in order
to derive the mass matrices for the quarks and leptons at the scale $\sim
10^{12}$ GeV. In this method one starts from the known and/or presumed known
quark and lepton masses and the mixing angles at the low energy scale and then
evolve these parameters, using the appropriate RGE's for the MSSM \cite{[BBO]},
to the desired scale  which in this case is situated at $10^{12} GeV$. After
evolving these parameters, the complex symmetric matrices M$^U$ and  M$^D$ for
the up and down are constructed with texture zeros \cite{[kus]}. For example
\begin{eqnarray} M_u=U_u^{\dag} D_uU_u=D_u+i\alpha
x[D_u,H]-{1\over2}\alpha^2x^2[[D_u,H],H]+...,
\end{eqnarray} and 
\begin{eqnarray} M_d=U_d^{\dag}D_dU_d=D_d+i\alpha
(x-1)[D_d,H]-{1\over2}\alpha^2(x-1)^2[[D_d,H],H]+...,
\end{eqnarray} where $D_u$ and $D_d$ are diagonal up and down mass matrices,
$\alpha$ is a real number and H is some hermitian matrix. Using Sylvester's
theorem the expression for
$\alpha H$ is given as:
\begin{eqnarray} i\alpha H=\sum^{3}_{k=1} ln(v_k){\Pi_{i\neq k}(V-v_i\times
1)\over
\Pi_{i\neq k}(v_k-v_i)}
\end{eqnarray} where V is the CKM matrix, and $v_i$'s are its eigenvalues which
are kept nondegenerate. The parameter x can be varied from 0 to 1 to obtain
different textures. The elements of those texures are then compared to remove
the lowest ones in order to get the textures with maximum numbers of zeros.
After constructing these textures, the masses and mixing angles are derived and
then are evolved to the weak scale to be compared with the experimental values.  

From our analysis described above, we have found only two types of textures (at
the partial unification scale), for the quark mass matrices, that are consistent
with the low energy data. Both type has at most four textue zeros.  Type 1 in
the symbolic form is:
\begin{eqnarray} M^{U}&=&\left(\matrix{
               a_u                &    &0     \cr
               0             &b_u         & 0 \cr
               0                &0        &c_u   }\right);
\end{eqnarray} and 
\begin{eqnarray} M^{D}&=&\left(\matrix{
               0                &a_d      &a_d^{\prime}     \cr
               a_d              &b_d         & b^\prime _d e^{i\beta} \cr
               a_d^{\prime}                &b^\prime _d e^{i\beta}        &c_d  
}\right)
\end{eqnarray}  Here all the parameters are real. The Type  2 looks like:
\begin{eqnarray} M^{U}&=&\left(\matrix{
               0                &a_u      &0     \cr
               a_u              &b_u         & b^\prime _u \cr
               0                &b^\prime _u        &c_u   }\right)
\end{eqnarray}      
\begin{eqnarray} M^{D}&=&\left(\matrix{
               0                &a_d      &0     \cr
               a_d              &b_d         & b^\prime _d e^{i\beta} \cr
               0                &b^\prime _d e^{i\beta}        &c_d   }\right)
\end{eqnarray}       For this type we also find the additional relation:
\begin{eqnarray}{b^\prime_u\over b_u}={b^\prime_d\over b_d}
\end{eqnarray} Consequently we have one less parameter in this type and obtain
one extra prediction. So  we will discuss this type 2 in the rest of the paper.

Now we will show analytically how the elements of this texture given by eqn.(6)
and eqn. (7), can be related to the masses and the CKM angles and whether there
is any further relation among the masses and the mixing angles. The elements of
the texture has a hierarchy, $c>>b\sim b^{\prime}>>a$.   Also the matrix $M^D$
can be re-written by removing the phases:
$M^D=P_DM^{\prime D}Q_D$, where P and Q are the phase matrices. To find the
quark masses at the scale $\sim$ 10$^{12}$ GeV, the matrices
$M^U$ and $M^{\prime D}$ need to be diagonalized. We use  the orthogonal
transformation
$RMR^{-1}=M^{\rm diag}$ to get the eigenvalues {$m_{u(d)},-m_{c(s)},m_{t,(b)}$}.
So the elements of the textures can be written in terms of the masses:
\begin{eqnarray} c_u\approx m_t(r); c_d\approx m_b(r); b_u\approx
-m_c(r);b_d\approx -m_s(r);a^2_u\approx m_u(r)m_c(r);a^2_d\approx m_d(r)m_s(r).
\end{eqnarray} where r is the partial unification scale, $10^{12}$ GeV.  From
Eqn. (8) and Eqn. (9), we obtain, 
\begin{eqnarray} {b^u\over b^d}={b^{\prime}_u\over
b^{\prime}_d}={m_c(r)\over m_s(r)} \end{eqnarray}
 The  matrix R looks like :
\begin{eqnarray} R&=&\left(\matrix{
               1                &s_1      &s_1s_2     \cr
               -s_1              &1         & s_2 \cr
               0                &-s_2        &1   }\right)
\end{eqnarray} For the up quark matrix, we obtain
\begin{eqnarray} s^u_1\equiv {\rm sin}\phi^u_1\simeq \sqrt{m_u\over m_c}
\end{eqnarray} and 
\begin{eqnarray}s^u_2\equiv {\rm sin}\phi^u_2\simeq {-b^\prime_u\over
c_u}\simeq {-b^\prime_u\over m_t}.
\end{eqnarray} For the down sector we have 
\begin{eqnarray}s^d_1\equiv {\rm sin}\phi^d_1\simeq \sqrt{m_d\over m_s}
\end{eqnarray} and 
\begin{eqnarray}s^d_2\equiv {\rm sin}\phi^d_2\simeq {-b^\prime_d\over
c_d}\simeq {-b^\prime_d\over m_b}.\end{eqnarray}
 Using Eqn.s (10), (13) and (15), we obtain
\begin{eqnarray} {s_2^d\over s_2^u}={m_t\over m_c}{m_s\over m_b}
\end{eqnarray}
 All the above equations are valid at the partial unification scale, r. So far,
other than
$b^{\prime}_u
$ and
$b^{\prime}_d$ and the phase $\beta$, all the other elements have been
determined in terms of the masses. Since
$b^{\prime}$'s are involved in the expressions for $s_2$'s we use the CKM matrix
elements. For the determination of phases we also use the CKM. At the scale $r$
$V_{\rm CKM}$ is given by:
\begin{eqnarray} V_{\rm CKM}(r)&=&R_u\left(\matrix{
               1                &      &   \cr
                             &e^{i\sigma}        & \cr
                               &       &e^{i\tau}   }\right)R_d^{-1},
\end{eqnarray}  
 where $\sigma=-2\beta$ and $\tau=-\beta$. Then, in terms the transforming angles
and the phases, the CKM element $V_{cb}(r)$ can be written as:
\begin{eqnarray}
\left|V_{cb}(r)\right|=\left|s^d_2e^{i\sigma}-s^u_2 e^{i\tau}\right|
\end{eqnarray} The eqn. (19) can be approximated as:
 \begin{eqnarray} \left|V_{cb}(r)\right|\approx{s^d_2} (1-2k)^{1/2}
\end{eqnarray} where 
\begin{eqnarray}k\equiv{m_c\over m_t}{m_b\over m_s}
\end{eqnarray} Using the magnitude of $V_{cb}(r)$, we can solve for $s^d_2$ ,
since $k$ is already known in terms the mass ratios at the intermediate
scale. For the determination of the phase we use the CKM element $V_{us}(r)$:
\begin{eqnarray}
\left|V_{us}(r)\right|=\left|-s^u_1+s^d_1e^{i\sigma}\right|
\end{eqnarray} Thus we have determined all the parameters of the model in terms
of the masses and some of the CKM elements.  The remaining CKM angles are the
predictions of the model. For example, the model predicts:
\begin{eqnarray}
\left|V_{ts}(r)\right|=\left|V_{cb}(r)\right|
\end{eqnarray} and 
\begin{eqnarray} V_{td}(r)=s^d_1 V_{cb}(r);\,\,\, V_{ub}(r)=s^u_1 V_{cb}(r)
\end{eqnarray} For the parametrization-invariant CP violation quantity J
\cite{[y]}, we obtain,
\begin{eqnarray} J&=&Im[V_{td}^*V_{tb}V_{ub}^*V_{ud}]\\\nonumber &\approx&s^u_1
s^d_1 {s^d_2}^2 {\rm Sin}2\beta,
\end{eqnarray} where J is determined at the weak scale. Now, we use the
experimental values of the quark masses and the CKM elements
$\left|V_{cb}\right|$ and
$\left|V_{us}\right|$ at the weak scale to determine the parameters of the
model. We use
$m_t=180 GeV$, and the light quark mass ratios \cite{[lw]} $ {m_u\over
m_d}=0.55$,
$ {m_s\over m_d}=18.8$ along with
$m_u(1 GeV)=5.1 MeV,m_d(1 GeV)=9.3 MeV,m_c=1.27 GeV,m_s(1 GeV)=0.175
GeV,m_b(m_b)=4.32 GeV, m_{\tau}(m_{\tau})=1.78 GeV$. The ratio of the values of
the Yukawa coupilng at the
$m_t$ scale to the 1 GeV or to the corresponding pole mass scale is given by
$\eta$.  The values of the
$\eta$ we use are
$\eta_u=2.4,\eta_d=2.4,\eta_s=2.4, \eta_c=2.1,\eta_{\tau}=1.0158$. The strong
coupling
$\alpha_s$ is taken to be 0.118  along with $sin^2\theta_w$ to be .2321 and
$\alpha$ =1/127.9 at the $M_Z$ scale. Also we use $\left|V_{cb}\right|=0.040$
and
$\left|V_{us}\right|=0.22$ at the $m_t$ scale. Now  we need to evolve these
masses and mixing elements to determine their values at the scale r. If we use a
large
$\tan\beta$($\tan\beta=48$) scenario where all the third generation Yukawa
coupling i.e top, bottom and $\tau$ are same, we get the masses as follows:
\begin{eqnarray} m_t(r)=158.4 GeV;m_c(r)=0.33 GeV; m_u(r)=0.001GeV;\\
m_d(r)=0.0019 GeV; m_s(r)=0.043 GeV; m_b(r)=3.30 GeV.
\end{eqnarray} At the  scale r, $\left|V_{cb}\right|$ becomes 0.03 and
$\left|V_{us}\right|$ is 0.22. We use these to solve for
$b^{\prime}$ and the phase $\beta$. We obtain 
\begin{eqnarray} s^d_2=0.0382;\, s^u_2=0.006;\, s^u_1=0.059;\, s^d_1=0.231\, \,
{\rm and}\, {\rm Sin}2\beta=0.96;
\end{eqnarray} Since the texture is already determined, we can determine the
other CKM parameters and J.  Using the above values, the predictions from the
model for the other CKM elements and J at the weak scale are:
\begin{eqnarray} V_{td}(m_t)=0.009;\, V_{ts}(m_t)=0.039;\, V_{ub}(m_t)=0.003;\,
J=3\times 10^{-5};
\end{eqnarray} One also can use a low $\tan\beta$ scenario just like the large
one. 

The superpotential for this texture can easily be written with some discrete
symmetry:\begin{eqnarray} W=\lambda_{33}F_3{\bar F_3}H_3 + ({s_1\over
M})\lambda_{32}{\bar F_3}F_2H_2+
 ({s\over M})^2\lambda_{22}{\bar F_2}F_2H_2+({s_1\over M})^3\lambda_{21}{\bar
F_2}F_1H_1 + h.c.
\end{eqnarray}  The fields $F_i$ and ${\bar F_i}$ correspond to the matter
superfield (2,1,4) and (1,2,${\bar 4}$). One can choose  $H_1$ and $H_3$ to be
the  bidoublet Higgs superfields and $H_2 $ to have the representation (2,2,15)
under the
$2_L2_R4_c$ symmetry in order to produce a
reasonable lepton mass matrix and  s's are the singlet  Higgs superfields. 
The quantum numbers for the field under the discrete symmetry we invoke can be as
follows:
\begin{eqnarray}  F_1:9;F_2:-11;F_3:2;s:5;s_1:-3;H_1:11;H_2:12;H_3:-4.
\end{eqnarray}$F_i$ and
${\bar F_i}$ transform in the same way under the discrete symmetry. There can be
other set of choices of the quantum numbers. However for this particular choice
of quantum numbers the 11 element of the up and down quark matrices get
generated for the 10th power of $s_1$, which can easily be neglected.   

In conclusion, we have derived two types of textures at the scale $\sim 10^{12}$
GeV which can give rise to the correct values for the masses of the quarks and
the leptons and the CKM angles and the CP violation parameter $J$. These
textures are different from those obtained at the GUT scale. Out of these two
types we found one has less number of parameters and some new relations among
the mixing angles and masses. This paticular texture also looks like the
Fritzsch texture with the ``22" element being non-zero. We also write a
superpotential for this texture supported by a model with string unification and
an intermediate scale around $10^{12}$ GeV.\\

One of us (SN) wishes to thank Duane Dicus of the University of Texas at Austin
for a very warm hospitality and support during this sabatical leave. This work
was supported in part by the US Department of Energy Grants No. DE-FG-02-94ER
40852 and DE-FG06-854ER 40224.

\end{document}